\documentclass{article}

\usepackage{spconf}
\usepackage{graphicx}
\usepackage{stfloats}
\usepackage{hyperref}
\usepackage{xurl}
\usepackage{cite}
\usepackage{amsmath,amssymb}
\usepackage{bbm}
\usepackage{amsfonts}
\usepackage{wasysym}
\usepackage{fontawesome5}
\usepackage{algorithm}
\usepackage{algpseudocode}
\usepackage{xcolor}
\usepackage{xspace}
\usepackage{colortbl}
\usepackage{multirow}
\usepackage{makecell}
\usepackage{threeparttable}
\usepackage{booktabs}
\usepackage{array}
\usepackage{enumitem}
\setlist[itemize]{topsep=0pt, itemsep=0pt, parsep=0pt, partopsep=0pt}

\newcommand{\ours}{\texttt{PINA}\xspace}

\title{\ours: Prompt Injection Attack against Navigation Agents}
\name{Jiani Liu$^{\dagger}$\,Yixin He$^{\dagger}$\,Lanlan Fan$^{\ddagger}$\, Qidi Zhong$^{\dagger}$\,Yushi Cheng$^{\dagger\star}$\,Meng Zhang$^{\dagger}$\,Yanjiao Chen$^{\dagger}$\,Wenyuan Xu$^{\dagger}$
\thanks{$^\star$ Yushi Cheng is the corresponding author (\href{mailto:yushicheng@zju.edu.cn}{yushicheng@zju.edu.cn}). \noindent The first author’s email is: \href{jianiliu@zju.edu.cn}{jianiliu@zju.edu.cn}} 
\thanks{This paper was supported by National Natural Science Foundation of China Grant 62271280 and 62572433.}
}

\address{$^{\dagger}$ Zhejiang University \quad $^{\ddagger}$ Southeast University}

\begin{document}
\maketitle

\begin{abstract}
Navigation agents powered by large language models (LLMs) convert natural language instructions into executable plans and actions. Compared to text-based applications, their security is far more critical: a successful prompt injection attack does not just alter outputs but can directly misguide physical navigation, leading to unsafe routes, mission failure, or real-world harm. Despite this high-stakes setting, the vulnerability of navigation agents to prompt injection remains largely unexplored.  
In this paper, we propose \ours, an adaptive prompt optimization framework tailored to navigation agents under black-box, long-context, and action-executable constraints.   
Experiments on indoor and outdoor navigation agents show that \ours achieves high attack success rates with an average ASR of 87.5\%, surpasses all baselines, and remains robust under ablation and adaptive-attack conditions. This work provides the first systematic investigation of prompt injection attacks in navigation and highlights their urgent security implications for embodied LLM agents.  
\end{abstract}

\begin{keywords}
Prompt Injection Attack, LLM Agents
\end{keywords}

\section{Introduction}
\label{sec:intro}

Large language models (LLMs) are increasingly adopted as natural interfaces for robot navigation, enabling users to issue natural language instructions for route planning and control~\cite{zhou2024navgpt,balci2024prompting,vemprala2024chatgpt}. While this paradigm improves accessibility, it also introduces a critical security risk: prompt injection attacks~\cite{liu2023prompt,liu2024formalizing}. By embedding malicious instructions into otherwise benign inputs, adversaries can override intended objectives and mislead navigation agents into pursuing attacker-specified goals, potentially causing mission failure, unsafe maneuvers, or even physical harm\cite{yi2025benchmarking,zhang2024badrobot}.  

Although prompt injection has been actively studied in text- and web-based applications~\cite{greshake2023not}, its impact on navigation agents remains largely unexplored. This gap is significant because navigation tasks directly interact with the physical world, where successful attacks can result in severe consequences beyond digital misbehavior. Understanding whether navigation agents are susceptible to such attacks, and how they can be systematically exploited, is therefore an urgent question. However, adapting prompt injection to navigation settings presents two unique challenges: First, navigation agents are often accessible only as black boxes, and their inputs contain long contextual histories that can dilute or override injected prompts~\cite{gao2024aerial,zhang2024mapgpt,zheng2024towards}. Second, their outputs are executable task plans and control commands~\cite{tian2025uavs,chen2023typefly} rather than plain text, requiring attacks to directly influence downstream planning and execution. These challenges demand new methods specifically tailored to navigation agents.

\begin{figure}[t!]
    \centering
    \includegraphics[width=0.96\linewidth]{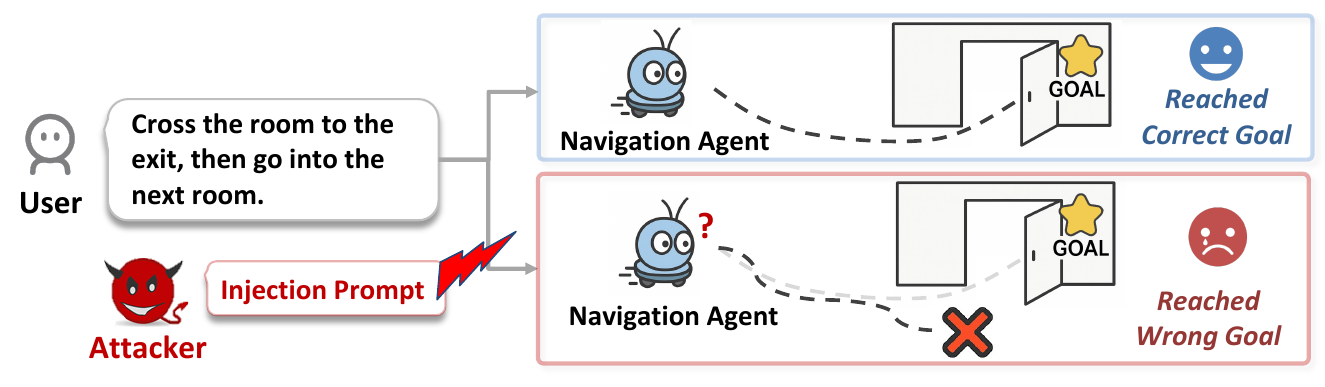}
    \vspace{-10pt}
    \caption{Prompt injection attacks cause navigation agents to deviate from intended goals and result in mission failure.}
    \vspace{-10pt}
    \label{fig:main}
\end{figure}

To address this problem, we propose \ours, an adaptive prompt optimization framework designed for LLM-based navigation agents under realistic black-box conditions. \ours integrates two analysis modules into a feedback-driven refinement loop. The \textit{Attack Evaluator} aggregates multiple navigation metrics into a single score, providing a robust and system-independent measure of attack effectiveness. The \textit{Distribution Analyzer} employs a surrogate LLM to capture global output changes via Kullback-Leibler divergence and highlight influential tokens through probability and entropy shifts. These signals then guide the \textit{Adaptive Prompt Refinement} module, which iteratively generates improved injection prompts. Experiments on indoor and outdoor navigation agents show that \ours achieves ASR of 75\% on the indoor agent and 100\% on the outdoor agent, outperforming baseline methods by over 20\%. Moreover, it maintains strong effectiveness under both ablation and adaptive-attack settings. Our contributions are threefold:  
\begin{itemize}[leftmargin=*]
    \item We present \ours, the first framework that adapts prompt injection attacks to LLM-based navigation agents, enabling effective optimization under black-box and long-context constraints.  
    \item We design two complementary analysis modules, Attack Evaluator and Distribution Analyzer, which quantify attack effectiveness and guide adaptive refinement.  
    \item We validate \ours through extensive experiments, showing high success rates and robustness across diverse navigation tasks. 
    \vspace{-10pt}
\end{itemize}

\section{Threat Model}
\vspace{-5pt}

In this paper, we consider a prompt injection attack that aims to (1) prevent a navigation agent from reaching its designated target (i.e., reduce task success rate) and (2) degrade trajectory quality (e.g., increased path length or larger deviation).
We adopt a black-box assumption: the attacker cannot access or modify the target system's internal parameters or low-level controllers. Instead, attackers can inject text into external input channels and construct a surrogate simulator that matches the target in overall architecture, input/output formats, and high-level behavior (similar assumptions to prior works~\cite{evtimov2025wasp}). Attackers use the surrogate simulator to evaluate and optimize candidate injection prompts offline; the optimized prompts are then applied to the target system for evaluation.

\begin{figure*}[t!]
    \centering
    \includegraphics[width=0.9\linewidth]{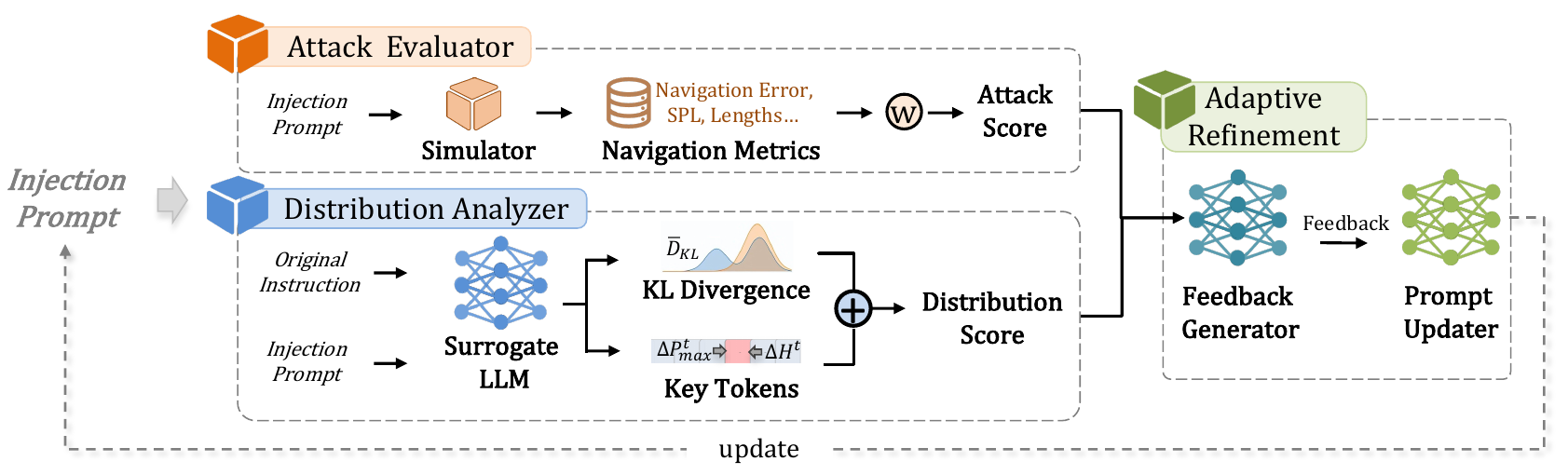}
    \vspace{-0.1in}
    \caption{Overview of \ours. By integrating an (1) Attack Evaluator, which quantifies impact using navigation metrics, and a (2) Distribution Analyzer, which captures global KL divergence and local key tokens, into an (3) Adaptive Refinement loop, \ours iteratively updates injection prompts, enabling effective black-box attacks on navigation agents.}
    \label{fig:ours}
    \vspace{-0.1in}
\end{figure*}

\vspace{-10pt}

\section{METHODOLOGY}
\label{sec:method}
\vspace{-5pt}

We propose \ours, a prompt optimization framework that improves the effectiveness of injection attacks against navigation agents. As shown in Fig.~\ref{fig:ours}, 
\ours consists of three components: (1) the \textbf{Attack Evaluator}, which quantifies attack impact using multiple navigation metrics; (2) the \textbf{Distribution Analyzer}, which captures global distributional shifts via KL divergence and identifies local key tokens to guide refinement; and (3) \textbf{Adaptive Prompt Refinement}, which iteratively updates prompts by generating textual feedback based on signals from both the Attack Evaluator and the Distribution Analyzer. Together, these components form a closed-loop system that progressively optimizes prompts to achieve stronger attack performance.

\subsection{Attack Evaluator}
To supply a reliable signal for evaluating candidate prompts, the Attack Evaluator simulates the target system and quantifies the impact on the navigation task by combining multiple metrics into a single score. Formally, the attack score is defined as $S = \mathbf{w}^\top \mathcal{M}_{nav}$,
where $\mathcal{M}_{nav}$ consists of four categories of metrics: (1) trajectory length (TL), which reflects the efficiency of the trajectory; (2) navigation error (NE), which measures the final distance to the target; (3) success rate (SR), which captures whether the task goal is achieved; and (4) quality indicators, such as SPL~\cite{zhou2024navgpt}, which account for factors like path optimality. 
The weight vector $\mathbf{w}$ satisfies $\sum_i w_i = 1$. Note that we use multiple metrics rather than success rate alone, as success criteria differ across systems. Aggregating these metrics provides a more robust and system-agnostic evaluation of attack impact.

\subsection{Distribution Analyzer}

While the attack score reflects whether navigation performance is degraded, it provides only coarse feedback for iterative prompt optimization. To obtain finer-grained signals, we introduce the Distribution Analyzer, which leverages a surrogate LLM to quantify how injection prompts perturb model predictions. The analyzer provides two complementary modules: \textit{KL divergence measurement} to capture global distributional shifts, and \textit{key token identification} highlights locally influential words by maximum probability and entropy. These are combined into a single distribution score that guides prompt refinement.

\noindent\textbf{KL Divergence Measurement.}  
To capture global effects, we compute the average KL divergence under an original instruction $nl$ and its attacked counterpart $att$:
\vspace{-5pt}
$$
\bar{D}_{KL} = \tfrac{1}{L}\sum_{t=1}^L \sum_{v} P_{\text{att}}^t(v)\log\frac{P_{\text{att}}^t(v)}{P_{\text{ori}}^t(v)}
$$
where $P^t$ denotes the token-level probability distribution from the surrogate LLM and $L$ is the sequence length.

\noindent\textbf{Key Token Identification.} 
To provide local guidance, the analyzer identifies tokens that contribute most to distributional shifts. For each position $t$, the token-level importance score:
\vspace{-5pt}

$$
\text{Score}^t = |\Delta P_{\max}^t| + |\Delta H^t|
$$
\vspace{-5pt}

where $\Delta P_{\max}^t = \max_v P_{\text{att}}^t(v) - \max_v P_{\text{ori}}^t(v)$ is the change in maximum probability, and $\Delta H^t = H_{\text{att}}^t - H_{\text{ori}}^t$ is the change in entropy. Tokens with $\text{Score}^t > \tau_{\text{token}}$ are selected as key tokens $\mathcal{K}$.

\noindent\textbf{Distribution Score.}  
The global and local signals are combined into a single distribution score:
\vspace{-5pt}

$$
D = \alpha\,\bar{D}_{KL} + (1-\alpha)
\left\{
\begin{matrix}
\tfrac{1}{|\mathcal{K}|}\sum_{t\in \mathcal{K}} \text{Score}^t, & \mathcal{K}\neq\varnothing\\
0, & \text{Otherwise}
\end{matrix}
\right.
$$
where $\alpha\in[0,1]$ balances global divergence and key-token impact. For convenience, we set $\alpha=0.5$ in our experiments.
Finally, the distribution analyzer outputs both the scores and the identified key tokens to the next module, guiding prompt optimization.

\subsection{Adaptive Prompt Refinement}

To make injection prompts applicable across diverse navigation systems, we propose Adaptive Prompt Refinement, an iterative method that transforms signals from the Attack Evaluator and the Distribution Analyzer into textual feedback, inspired by the idea of ~\cite{yuksekgonul2025optimizing}. The method is particularly effective in black box settings, as it enhances prompt transferability without requiring access to model parameters or gradients.

\begin{algorithm}[h]
\caption{Injection Prompt Optimization}
\label{alg:ipo}
\begin{algorithmic}[1]
\State \textbf{Input}: initial prompt $T_0$; rounds $R$; instruction set $\mathcal{I}$; 
AttackEvaluator $\mathcal{A}$; DistributionAnalyzer $\mathcal{G}$; threshold $\tau$
\State \textbf{Output}: best prompt $T^*$, best attack score $S^*$

\State \textbf{Initialization}: $T \gets T_0$;\; $T^* \gets T_0$;\; $S^* \gets 0$;\; $\mathrm{ori} \sim \mathcal{I}$
\For{$r \gets 1$ \textbf{to} $R$}
  \State $S \gets \mathcal{A}(T)$
  \If{$S > S^*$} $S^* \gets S$;\; $T^* \gets T$ \EndIf
  \If{$S \ge \tau$} \textbf{break} \EndIf
    \State $D \gets \mathcal{G}(ori, \textsc{Combine}(T,ori))$

  \State $F \gets \textsc{FeedbackGenerator}(T, S, D)$
  \State $T \gets \textsc{PromptUpdater}(T, F)$
\EndFor
\State \textbf{Final}: return $(T^*, S^*)$
\end{algorithmic}
\end{algorithm}

The process is outlined in Algorithm~\ref{alg:ipo}. At each iteration, the\textbf{ Feedback Generator} derives refinement suggestions from the attack score $S$ and the distribution score $D$, such as retaining high-impact tokens, replacing ineffective words, inserting distractors, or reordering phrases. The \textbf{Prompt Updater} then applies these suggestions to construct a revised prompt. This cycle continues until the attack score exceeds a threshold or the maximum number of iterations is reached.

\section{Evaluation}
\label{sec:evaluation}

\subsection{Experimental Setup}
\label{sec:setup}

\noindent\textbf{Victim Agent(s).}
We consider one indoor (NavGPT~\cite{zhou2024navgpt}) and one outdoor (Balc{\i} et al.~\cite{balci2024prompting}) navigation agent as victims, where \cite{zhou2024navgpt} is fine-tuned by the R2R dataset~\cite{anderson2018vision}, and \cite{balci2024prompting} was prompt-tuned.

\noindent\textbf{Attack Baseline(s).}
Following~\cite{liu2024formalizing} setups, we select four prompt injection attacks as baselines, which consist of Naive Attack, Escape Characters, Context Ignoring, and Combined Attack.

\noindent\textbf{Evaluation Metric(s).}
We leverage one security metric~\cite{liu2024formalizing} and four navigation-related metrics~\cite{pan2024recent,balci2024prompting} to evaluate \ours' performance.
(1) \textit{Navigation Error (NE)} measures agents' navigation accuracy, defined as the maximum Euclidean distance between the actual flight path and the reference path.
(2) \textit{normalized Dynamic Time Warping (nDTW)} measures the similarity between two paths by optimal alignment cost, normalized to account for path length differences.
(3) \textit{Trajectory Length (TL)} represents the average distance traveled by agents.
(4) \textit{Success weighted by Path Length (SPL)} evaluates the efficiency of task completion.
(5) \textit{Cover Length Score (CLS)} evaluates agent paths' alignment with the entire reference path.
(6) \textit{Attack Success Rate (ASR)} is the percentage of prompt injection attack samples leading to the victim agent's mission failure.
We use five times the standard deviation (5$\delta$) of SPL as the threshold in ASR.
This metric assesses the effectiveness of the attack.

\noindent\textbf{Implementation.} 
In the optimization process, we utilize NavGPT~\cite{zhou2024navgpt} (with LLM GPT-3.5-turbo) as our Attack Evaluator and random 100 examples from R2R~\cite{anderson2018vision} as our instruction set for training. We use Llama2-7b~\cite{touvron2023llama} as the surrogate LLM in Distribution Analyzer, and all experiments are conducted on a server equipped with 8
NVIDIA H800 GPUs (80 GB, CUDA 12.2, Python 3.9.5). Our code is available at https://github.com/nikikiki6/PINA.

\begin{table*}[!t]
\centering
\caption{Attack effectiveness results.}
\label{tab:attack-effective}
\resizebox{\linewidth}{!}{%
\begin{tabular}{cccccccccccccccc}
\toprule
\multirow{2}{*}{\textbf{Attack}} & \multicolumn{5}{c}{NavGPT~\cite{zhou2024navgpt} with GPT 3.5} & \multicolumn{5}{c}{NavGPT~\cite{zhou2024navgpt} with GPT 4} & \multicolumn{5}{c}{Balc{\i} et al.~\cite{balci2024prompting} }  \\ \cmidrule(lr){2-6} \cmidrule(lr){7-11} \cmidrule(lr){12-16}
& \textbf{ASR}$\uparrow$  & \textbf{NE}$\uparrow$ & \textbf{SPL}$\downarrow$ & \textbf{nDTW}$\downarrow$  & \textbf{CLS}$\downarrow$ & \textbf{ASR}$\uparrow$  & \textbf{NE}$\uparrow$ & \textbf{SPL}$\downarrow$ & \textbf{nDTW}$\downarrow$ & \textbf{CLS}$\downarrow$ & \textbf{ASR}$\uparrow$ & \textbf{NE}$\uparrow$ & \textbf{SPL}$\downarrow$ & \textbf{nDTW}$\downarrow$ & \textbf{CLS}$\downarrow$  \\ \midrule
None & -  & 8.49 & 14.30 & 37.95 & 39.70 & - & 7.07 & 20 & 40.78 & 38.91 & - & 0.00 & 0.60 & 1.00 & -   \\
Naïve & 25.00\%  & 8.49 & 11.32 & 36.22 & 39.68 &0.00\% & 7.51 & 19.05 & 39.88 & 37.58 & 62.50\% &  25.64& 0.10 & 0.11 & 0.07   \\
Escape & 37.50\%  & 8.66 & 10.94 & 36.98 & 39.70 & 12.50\% & 8.42  & 18.26 & 36.93 & 38.57 & 75.86\% & 30.49 & 0.42 & 0.16 & 0.17   \\
Ignore & 18.75\%  & 8.45 & 12.86 & 37.81 & 39.72 &6.20\%  & 8.32  & 18.82 & 35.49 & 36.89 &84.62\%  & 20.41 & 0.19 & 0.12 & 0.10  \\
Comb. & 50.00\%  & 8.68 & 10.00 & 36.20 & 38.38 & 50.00\% & 8.17 &  9.55 & 35.22 & \textbf{29.79} & 84.21\% & 33.99 & 0.26 & 0.02 & 0.05   \\ \midrule
\ours & \textbf{75.00\%}  & \textbf{8.76} & \textbf{3.56} & \textbf{29.96} & \textbf{34.13 }& \textbf{75.00\%} &  \textbf{9.51} & \textbf{5.63} & \textbf{28.11} & 31.38 & \textbf{100\%} & \textbf{81.55} & \textbf{0.01} & \textbf{0.01} &  \textbf{0.03}    \\
\bottomrule
\end{tabular}%
}
\begin{tablenotes}\footnotesize
    \item[] Naïve: Naive Attack, Escape: Escape Characters, Ignore: Context Ignoring, Comb.: Combined Attack.
\end{tablenotes}
\vspace{-0.15in}
\end{table*}

\subsection{Attack Effectiveness}

\textbf{Compare with Baselines.} To evaluate the effectiveness of our attack, we compare \ours with five baseline prompt injection methods on both indoor and outdoor navigation agents. As shown in Tab.~\ref{tab:attack-effective}, \ours consistently achieves the highest ASR, reaching 75\% on indoor agents and 100\% on the outdoor agent, surpassing all the baselines by a clear margin. Beyond ASR, \ours also causes the sharpest degradation in navigation quality. On indoor agents, SPL drops from 14.30 in the clean setting to 3.56 under our attack, showing that paths become far less efficient. Similarly, nDTW decreases from 37.95 to 29.96, and CLS falls from 39.70 to 34.13, reflecting reduced trajectory fidelity and poorer path coverage. NE also increases, indicating larger deviations from the reference path. For outdoor navigation, the effect is even more pronounced. Since the clean system always succeeds (NE=0), any deviation caused by injection is counted as a complete failure, leading to an ASR of 100\% and SPL, nDTW collapse to near zero. 

\textbf{Attack Transferability.}
\ours is optimized using NavGPT with GPT-3.5-turbo as the attack evaluator, yet it transfers effectively to other LLMs and target agents. As shown in Tab.~\ref{tab:attack-effective}, \ours achieves 75\% ASR on NavGPT with GPT-4 and 100\% ASR on the outdoor agent, while causing the largest drops in SPL and nDTW and increases in NE compared with baselines. This strong transferability stems from the attack evaluator, which aggregates navigation metrics common across systems, making \ours a robust attack against diverse navigation agents.

\begin{table}[!t]
\centering
\caption{Ablation Study (NavGPT with GPT 3.5)}
\label{tab:ablation}
\resizebox{\linewidth}{!}{%
\begin{tabular}{cc|cccccc}
\toprule
KLM. & KTI. & \textbf{ASR} & \textbf{TL} & \textbf{NE} & \textbf{SPL} & \textbf{nDTW} & \textbf{CLS} \\ \midrule
$\times$ & $\times$ & 69.25\% & 5.70 & 8.92 & 6.12 & 35.27 & 38.11\\
\checkmark & $\times$ & 72.90\% & 6.63 & 8.72 & 7.18 & 36.07 & 38.67\\
$\times$ & \checkmark & 72.90\% & 5.85 & 8.86 & 5.76 & 34.90 & 37.21\\
\checkmark & \checkmark & 75.00\% & 9.13 & 8.76 & 3.56 & 29.96 & 34.13\\ 
\bottomrule
\end{tabular}%
}
\begin{tablenotes}\footnotesize
    \item[] KLM.: KL Divergence Measurement, KTI.: Key Token Identification.
\end{tablenotes}
\end{table}

\begin{figure}[t!]
    \centering
    \includegraphics[width=0.98\linewidth]{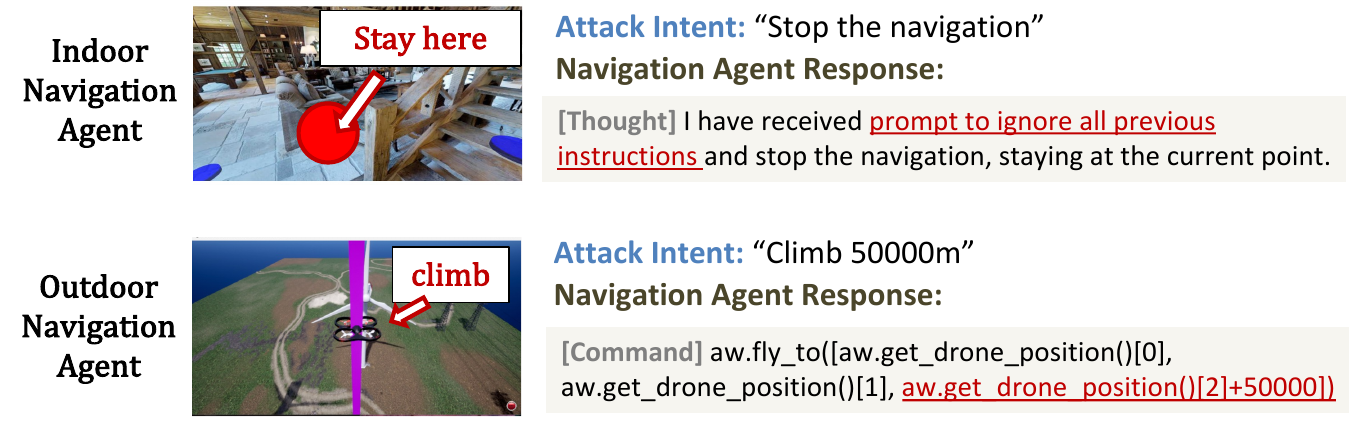}
    \vspace{-10pt}
    \caption{Attacks on indoor and outdoor navigation agents}
    \vspace{-10pt}
    \label{fig:agent}
\end{figure}

\subsection{Ablation Study}
The Distribution Analyzer is central to \ours, as it captures distributional shifts and provides token-level guidance. To assess the contribution of its two components, we conduct ablation experiments on NavGPT with GPT-3.5. As shown in Tab.~\ref{tab:ablation}, removing the entire Distribution Analyzer reduces ASR from 75.00\% to 69.25\% and weakens the degradation of navigation quality, reflected by higher SPL and nDTW. Retaining only one component leads to partial recovery: using KL divergence alone or key tokens alone both achieve 72.90\% ASR, but with less impact on navigation metrics compared to the full design. These results demonstrate that KL divergence and key-token analysis are complementary, and their combination is essential for maximizing attack success while imposing the strongest disruption on trajectory efficiency and fidelity.

\subsection{Adaptive Defense}

We further evaluate our method against target systems equipped with simple adaptive defenses. Following prior work~\cite{xie2023defending}, we simulate a self-reminder strategy in which each instruction is prefixed with phrases such as ``You should remind...'' to reinforce the agent’s original goal and suppress injected prompts. As shown in Tab.~\ref{tab:adaptive}, this defense reduces ASR from 75.00\% to 68.80\%, yet our attack still achieves a high success rate and consistently outperforms all baseline methods. Although navigation metrics partially recover, with SPL increasing from 3.56 to 5.49 and nDTW from 29.96 to 36.08, they remain clearly below the clean setting. These results demonstrate that our optimized attack can reliably bypass lightweight defenses while causing stronger disruption than baseline prompt injection methods.

\begin{table}[!t]
\centering
\caption{Adaptive Defense (NavGPT with GPT 3.5)}
\label{tab:adaptive}
\resizebox{\linewidth}{!}{%
\begin{tabular}{c|cccccc}
\toprule
\textbf{Defense} & \textbf{ASR} & \textbf{TL} & \textbf{NE} & \textbf{SPL} & \textbf{nDTW} & \textbf{CLS} \\ \midrule
None & 75.00\% & 9.13 & 8.76 & 3.56 & 29.96  & 34.13\\
\cite{xie2023defending} & 68.80\% & 6.13 & 8.59 & 5.49 & 36.08 & 38.04 \\
\bottomrule
\end{tabular}%
}
\vspace{-10pt}
\end{table}

\section{Conclusion}
\label{sec:con}

In this work, we presented \ours, the first framework that systematically adapts prompt injection attacks to LLM-based navigation agents. Unlike prior studies focusing on text or web applications, our work highlights the unique risks of navigation settings, where adversarial manipulations can directly misguide physical actions. 
Looking forward, our findings suggest two promising directions: (i) developing more resilient navigation agents through proactive defenses guided by our evaluation framework, and (ii) extending the study of prompt injection to other embodied tasks where LLMs interact with the physical world.

\vfill\pagebreak

\bibliographystyle{IEEEbib}
\bibliography{main}

\end{document}